\documentclass{ifacconf}

\usepackage{amsfonts,amsmath} % From the American Mathematical Society
\usepackage{graphicx}         % include this line if your document contains figures
\usepackage{natbib}           % required for bibliography

\newcommand{\E}[1]{\mathop{{\rm \bf E}\left\{#1\right\}}\nolimits}

\usepackage{clrscode}

\begin{document}
\begin{frontmatter}

\title{On the Stable Cholesky Factorization-based Method for the Maximum Correntropy Criterion Kalman Filtering\thanksref{footnoteinfo}}
% Title, preferably not more than 10 words.

\thanks[footnoteinfo]{The author acknowledges the financial support of the Portuguese FCT~--- \emph{Funda\c{c}\~ao para a Ci\^encia e a Tecnologia}, through the projects UIDB/04621/2020 and UIDP/04621/2020 of CEMAT/IST-ID, Center for Computational and Stochastic Mathematics, Instituto Superior T\'ecnico, University of Lisbon.}

\author[First]{Maria V. Kulikova}
%\author[Second]{Julia V. Tsyganova}

\address[First]{CEMAT (Center for Computational and Stochastic Mathematics), Instituto Superior T\'{e}cnico, Universidade de Lisboa, Av. Rovisco Pais 1,  1049-001 LISBOA, Portugal (email: maria.kulikova@ist.utl.pt).}
%\address[Second]{Department of Mathematics, Information and Aviation Technology, Ulyanovsk State University, Str. L. Tolstoy 42, 432017 Ulyanovsk, Russian %Federation (email: TsyganovaJV@gmail.com).}

\begin{abstract}                % Abstract of not more than 250 words.
This paper continues the research devoted to the design of  numerically stable square-root implementations for the maximum correntropy criterion Kalman filtering (MCC-KF). In contrast to the previously obtained results, here we reveal the first robust (with respect to round-off errors) method within the  Cholesky factorization-based approach. The method is formulated in terms of square-root factors of the {\it covariance} matrices, i.e. it belongs to the covariance-type filtering methodology. Additionally, a  numerically stable orthogonal transformation is utilized at each iterate of the algorithm for accurate propagation of the Cholesky factors involved. The results of numerical experiments illustrate a superior performance of the novel MCC-KF implementation compared to both the conventional algorithm and its previously published Cholesky-based variant.
\end{abstract}


\end{frontmatter}
%===============================================================================

\section{Introduction}
\label{Introduction}

One of the most recent solution to the problem of ``distributionally'' robust filtering (i.e. when the actual distribution deviates from the ``nominal'' one) is obtained under the so-called maximum correntropy criterion (MCC). More precisely, when the classical state-space models are examined, the  ``nominal'' distribution is assumed to be Gaussian, and the goal is to enhance the underlying filter performance in terms of estimation accuracies in a case of the presence of outliers. The obtained MCC-based estimators with the Kalman filtering (KF) like structure (i.e. the first two moments are computed, only) are derived for both the linear stochastic systems in~\cite{2007:Liu,2014:Chen,2015:Chen,2011:Cinar,2012:Cinar,2016:Izanloo,2017:Chen,2017:Liu:KF,2017:CSL:Kulikova,2019:Fakoorian} and the nonlinear state-space models in~\cite{2016:Liu:EKF,2018:Kulikov:SP,2017:Liu:UKF,2017:Qin,2017:Wang,2016:Wang,2020:IFAC:Kulikov}.  One of the obtained estimators is called the maximum correntropy criterion Kalman filter (MCC-KF) as proposed in~\cite{2016:Izanloo}. This estimation method is widely used for solving practical problems; e.g., see~\cite{2017:Yang,2018:Yang} and many other studies.

Recently, the numerical robustness issues of the mentioned MCC-KF estimator have been investigated in~\cite{2019:SP:Kulikova}. The key problem in this research area is to derive a stable (in a finite precision arithmetic) square-root MCC-KF implementation methods, which are demanded for solving applications with high reliability requirements as discussed in~\cite{2010:Grewal,2019:Grewal} and many other works. In this paper, we focus on the traditional square-root strategy that is based on the Cholesky factorization applied to covariance matrices involved in the filter; see~\cite{2000:Kailath:book,2006:Simon:book,2015:Grewal:book}. Unfortunately, the previously discovered Cholesky-based MCC-KF method in~\cite{2019:SP:Kulikova} is shown to possess a poor performance in ill-conditioned state estimation scenario. Thus, the goal of this research is to answer a question whether it is possible or not to find a more reliable MCC-KF implementation within the discussed class of Cholesky factorization-based algorithms. We answer positively this question and derive the first stable square-root MCC-KF method that outperforms for estimation accuracy both the conventional MCC-KF algorithm and its previously published Cholesky-based variant.

More precisely, the novel estimator is designed in terms of the {\it lower} triangular Cholesky factors of the error {\it covariance matrices}, i.e. it is of covariance-type filtering methodology. Additionally, stable orthogonal rotations are utilized as far as possible for propagating the involved Cholesky factors.  The performance of the examined MCC-KF implementation methods are studied by using a sixth-order radar tracking system example.

\section{PROBLEM STATEMENT}\label{problem:statement}

Consider a linear discrete-time state-space model
  \begin{align}
   x_{k} = & F_{k-1} x_{k-1}  + G_{k-1} w_{k-1},  \quad k \ge 1, \label{eq:st:1} \\
    y_k  = & H_k x_k + v_k   \label{eq:st:2}
  \end{align}
where the system matrices $F_k \in \mathbb R^{n\times n}$, $G_k \in \mathbb R^{n\times q}$, $H_k \in \mathbb R^{m\times n}$ and the noises' covariances $Q_k \in \mathbb R^{q\times q}$ ($Q_k > 0$) and $R_k \in \mathbb R^{m\times m}$ ($R_k \ge 0$) are known. The vectors $x_k \in \mathbb R^n$  and $y_k \in \mathbb R^m$ are, respectively, the hidden dynamic state to be estimated and the available measurement vector. The random variables $x_0$, $w_k$ and $v_k$ are assumed to satisfy
\begin{align*}
&\E{x_0}       = \bar x_0,    &  &\E{(x_0-\bar x_0)(x_0-\bar x_0)^{\top}}  = \Pi_0, \\
&\E{w_k}       = \E{v_k} = 0, &  &\E{w_kx_0^{\top}} = \E{v_kx_0^{\top}}  = 0,  \\
&\E{w_kv_k^{\top}}  = 0,           &  &\E{w_kw_j^{\top}} = Q_k\delta_{kj}, \\
&                             &  &\E{v_kv_j^{\top}} = R_k\delta_{kj}
\end{align*}
where the symbol $\delta_{kj}$ is the Kronecker delta function, and the initial mean $\bar x_0$ and error covariance $\Pi_0 \ge 0$ are known.

The minimum {\it linear} expected mean square error (MSE) estimator derived for the examined state-space model~\eqref{eq:st:1}, \eqref{eq:st:2} is known as the Kalman filter (KF); see Theorem~9.2.1 in~\cite{2000:Kailath:book}. In case of Gaussian uncertainties in the model, the minimum expected MSE estimate belongs to a class of linear functions and, hence, being a linear estimator the KF provides the optimal estimate in the MSE sense; e.g., see~\cite{2017:Aravkin}. However, in non-Gaussian settings, the classical KF exhibits only sub-optimal behavior under the minimum expected MSE estimation criterion.

The concept of {\it correntropy} (that is a similarity measure of two random variables) has become a very popular technique in the past few years in the realm of designing the ``distributionally'' robust estimators, i.e. when the actual distribution deviates from the ``nominal'' one. For the examined state-space models, the  ``nominal'' distribution is assumed to be Gaussian, and the goal is to enhance the classical KF performance in a case of outliers appearance. One of the resulted estimators is called the maximum correntropy criterion Kalman filter (MCC-KF) as proposed in~\cite{2016:Izanloo}. In general, the maximum correntropy cost function is used in the related estimation problem as follows, e.g., see the details in Chapter~5 in~\cite{2018:Principe:book}: an estimator of unknown state $X \in {\mathbb R}$ can be defined as a function of observations $Y \in {\mathbb R}^m$, i.e. $\hat X = g(Y)$ where $g$ is solved by maximizing the correntropy between $X$ and $\hat X$, i.e.
\begin{equation} \label{mcc:kriterion}
 \mbox{arg}\max \limits_{g \in G} V(X,\hat X) = \mbox{arg}\max \limits_{g \in G} \E{k_{\sigma}\Bigl(X - g(Y)\Bigr)}
\end{equation}
where $G$ stands for the collection of all measurable functions of $Y$, $k_{\sigma}(\cdot)$ is a kernel function and  $\sigma > 0$ is the kernel size (bandwidth). As mentioned above, the ``nominal'' distribution is assumed to be Gaussian and, hence, we explore the Gaussian kernel in the related estimation problem given as follows:
\begin{equation}\label{Gauss_kernel}
k_{\sigma}(X - \hat X) = \exp \left\{ -{(X - \hat X)^2}/{(2\sigma^2)}\right\}.
\end{equation}
It is not difficult to see that the MCC cost~\eqref{mcc:kriterion} with kernel~\eqref{Gauss_kernel} reaches its maximum if and only if $X = \hat X$.

In summary, the discussed MCC-KF technique for estimating the unknown dynamic state $\hat x_{k|k}$ in the state-space model~\eqref{eq:st:1}, \eqref{eq:st:2} with the Gaussian kernel $k_{\sigma}(\cdot)$ from~\eqref{Gauss_kernel} is derived by maximizing the following cost function:
\begin{align}
 J(k) & = k_{\sigma}(\|\hat x_{k|k}-F_{k-1}\hat x_{k-1|k-1}\|_{P_{k|k-1}^{-1}})  \nonumber \\
 & + k_{\sigma}(\|y_k-H_k\hat x_{k|k}\|_{R_k^{-1}}) \label{MCCKF:cost}.
\end{align}

The stated optimization problem has been solved by~\cite{2012:Cinar,2016:Izanloo} where the arisen nonlinear equation was resolved with respect to $\hat x_{k|k}$ by utilizing a fixed point rule with one iterate, only. More precisely, this approach yields the following filtering recursion:
\begin{equation}
\hat x_{k|k}  =  F_{k-1}\hat x_{k-1|k-1}  + K_k(y_k - H_k \hat x_{k|k-1}) \label{eq:result}
\end{equation}
where the gain matrix is computed by
\begin{equation}
K_{k}  = \lambda_k \left( P_{k|k-1}^{-1}+\lambda_k H_k^{\top} R_k^{-1} H_k \right)^{-1}H_k^{\top} R_k^{-1} \label{eq:gain}
\end{equation}
and $\lambda_k$ is a scalar adjusting weight as suggested in~\cite{2016:Izanloo}
\begin{equation}
\lambda_{k}  = \frac{k_{\sigma}(\|y_k-H_k\hat x_{k|k-1}\|_{R_k^{-1}})}{k_{\sigma}(\|\hat x_{k|k-1}-F_{k-1}\hat x_{k-1|k-1}\|_{P_{k|k-1}^{-1}})}.  \label{eq:lambda}
\end{equation}
Finally, the recursion for the state estimate in~\eqref{eq:result} is combined with the symmetric Joseph stabilized formula for calculating the filter error covariance matrix, which has been derived for the classical KF; e.g., see~\cite{2000:Kailath:book,2006:Simon:book,2015:Grewal:book}. The pseudo-code in Algorithm~1  summarizes the discussed MCC-KF suggested in~\cite{2016:Izanloo}.

%_______________________________________
\begin{codebox}
\Procname{{\bf Algorithm 1}. $\proc{MCC-KF}$ ({\it conventional MCC-KF})}
\zi \textsc{Initialization:}($k=0$) $\hat x_{0|0} = \bar x_0$ and $P_{0|0} = \Pi_0$.
\zi \textsc{Time Update}: ($k=\overline{1,N}$)
\li \>\!\!$\hat x_{k|k-1}  = F_{k-1} \hat x_{k-1|k-1}$; \label{mcc:p:X}
\li \>\!\!$P_{k|k-1}  = F_{k-1} P_{k-1|k-1}F_{k-1}^{\top}+G_{k-1}Q_{k-1}G_{k-1}^{\top}$; \label{mcc:p:P}
\zi \textsc{Measurement Update}: ($k=\overline{1,N}$)
\li \>\!\!Compute $\lambda_k$ by formula~\eqref{eq:lambda};  \label{mcc:f:L}
%$\lambda_{k}  = \displaystyle\frac{G_{\sigma}\left(\|y_k-H_k \hat x_{k|k-1}\|_{R^{-1}_k}\right) }{G_{\sigma}\left(\|\hat x_{k|k-1}-F_{k-1} \hat %x_{k-1|k-1}\|_{P_{k|k-1}^{-1}}\right)}$;
\li \>\!\!$K_{k}  = \lambda_k \left( P_{k|k-1}^{-1}+\lambda_k H_k^{\top} R_k^{-1} H_k \right)^{-1}H_k^{\top} R_k^{-1}$; \label{mcc:f:K}
\li \>\!\!$P_{k|k}  = (I- K_{k}H_k)P_{k|k-1}(I- K_{k}H_k)^{\top}+K_k R_k K_k^{\top}$;  \label{mcc:f:P}
\li \>\!\!$\hat x_{k|k}  =    \hat x_{k|k-1}+K_{k}(y_k-H_k \hat x_{k|k-1})$.   \label{mcc:f:X}
\end{codebox}
%_______________________________________

Algorithm~1 is said to be of the {\it covariance} type presented in the {\it conventional} form, because it recursively processes the {\it full} error {\it covariance} matrices $P_{k|k-1}$ and $P_{k|k}$ at each iterate of the filter. Taking into account that any covariance matrix is a symmetric matrix, it makes sense to propagate only half of them. The traditional way utilized in the KF community is the use of Cholesky factorization. It implies the decomposition $P = SS^{\top}$ and, then, the underlying filtering recursion is re-derived in terms of propagating the Cholesky factor $S$, only; see~\cite{1975:Morf,1995:Park,1977:Bierman} and many other studies. This computational approach is also known to improve the numerical stability of any conventional KF-like implementation in a finite precision arithmetic, because it ensures the symmetric form and positive (semi-) definiteness of the original matrix $P$ (while the recovering by backward multiplication $SS^{\top}=P$) despite the influence of roundoff errors; see~\cite{2000:Kailath:book,2006:Simon:book,2015:Grewal:book}. Recently, the Cholesky factorization-based method has been derived in~\cite{2019:SP:Kulikova}. However, the numerical stability of the suggested method is still poor as illustrated by the results of numerical experiments presented in the cited paper. The question to be answered in this paper is the following: whether it is possible or not to find a more reliable MCC-KF implementation within the discussed Cholesky-based class of methods.

\section{MCC-KF CHOLESKY-BASED FILTERING}\label{main:result}

We start our research with the MCC-KF square-root method previously designed in~\cite{2019:SP:Kulikova}. The goal is to improve the method in terms of its stability with respect to roundoff errors. It is worth noting here that the first Cholesky factorization-based MCC-KF implementation derived in the cited paper is formulated in terms of the upper triangular Cholesky factors. For readers' convenience, we first re-formulate it in terms of the lower triangular matrices and, next, we propose its robust alternative. More precisely, let's consider the Cholesky factorization of a symmetric positive definite matrix $A$ in the form $A=A^{1/2}A^{{\top}/2}$ where the factor $A^{1/2}$ is a {\it lower} triangular matrix with positive diagonal entries. The square-root MCC-KF algorithms imply the following  essential features: (i) the factorization is only performed for $\Pi_0>0$; (ii) the filtering equation in Algorithm~1 are then re-derived for propagating the lower triangular matrices $P_{k|k-1}^{1/2}$ and $P_{k|k}^{1/2}$; (iii) numerically stable orthogonal rotations are utilized for updating the involved Cholesky factors that additionally provide the {\it array} form suitable for parallel implementation. Finally, it is important that the adjusting weight $\lambda_k$ defined in~\eqref{eq:lambda} is a nonnegative value and, hence, a square root exists.

To design the mathematically equivalent analogue of Algorithm~1, we factorize its equations as follows:
\begin{align*}
&P_{k|k-1} = F_{k-1} P_{k-1|k-1}F_{k-1}^{\top}+G_{k-1}Q_{k-1}G_{k-1}^T = AA^{\top}  \\
&\!\!=[F_{k-1}\!P_{k-1|k-1}^{1/2}, G_{k-1}Q_{k-1}^{1/2}][F_{k-1}\!P_{k-1|k-1}^{1/2}, G_{k-1}Q_{k-1}^{1/2}]^{\top}
\end{align*}
where an orthogonal rotation, say $Q$, is applied to get the corresponding lower triangular post-array, $R$, by transformation $AQ=R$, i.e. we have
\[ RR^{\top} = (AQ)(AQ)^{\top} = AA^{\top}  = [X, \; 0][X, \; 0]^{\top} = P_{k|k-1} \]
and, hence, we conclude that $X:=P_{k|k-1}^{1/2}$ is the resulted square-root factor, which we are looking for. This value is pulled out from the post-array, if required.

Next, the following formulas have been proved for the filter gain $K_k$ calculation in~\cite{2017:CSL:Kulikova}:
\begin{align}
K_k & = \lambda_k \hat P_{k|k}H_k^{\top}R_k^{-1} \nonumber \\
& = \lambda_k\left( P_{k|k-1}^{-1}+\lambda_kH_k^{\top}R_k^{-1}H_k\right)^{-1}H_k^{\top}R_k^{-1}.  \label{mcc:K:eq2} \\
K_{k} & = \lambda_k P_{k|k-1}H_k^{\top}R_{e,k}^{-1} \nonumber \\
& = \lambda_k P_{k|k-1}H_k^{\top}(\lambda_kH_kP_{k|k-1}H_k^{\top}+R_k)^{-1}. \label{mcc:K:eq1}
\end{align}

It is not difficult to see that the MCC-KF estimator involves equation~\eqref{mcc:K:eq2} for computing $K_k$ in line~\ref{mcc:f:K} of Algorithm~1. To derive the square-root form, we consider the term in the brackets, i.e. $\hat P_{k|k}^{-1} = P_{k|k-1}^{-1}+\lambda_kH_k^{\top}R_k^{-1}H_k$, and factorize it in the form $AA^{\top}$ where the post array is  obtained by transformation $R = AQ$. Thus, we get
\begin{align*}
AA^{\top} & = P_{k|k-1}^{-1}+\lambda_kH_k^{\top}R_k^{-1}H_k \\
& =[P_{k|k-1}^{-{\top}/2}, \; \lambda_k^{1/2}H_k^{\top}R_k^{-{\top}/2}][P_{k|k-1}^{-{\top}/2}, \; \lambda_k^{1/2}H_k^{\top}R_k^{-{\top}/2}]^{\top} \\
& = RR^{\top}  =\hat P_{k|k}^{-1} = [\hat P_{k|k}^{-{\top}/2}, \; 0][\hat P_{k|k}^{-{\top}/2}, \; 0]^{\top}.
\end{align*}

When the resulted block $[\hat  P_{k|k}^{-{\top}/2}]$ is read-off from the post-array, the gain matrix is calculated as follows:
\[K_{k}  = \lambda_k P_{k|k}H_k^{\top}R_k^{-1}  = \lambda_k\Bigl([\hat P_{k|k}^{-{\top}/2}][\hat P_{k|k}^{-{\top}/2}]^{\top}\Bigr)^{-1}H_k^{\top} R_k^{-1}. \]

Finally, although the Cholesky factor $\hat P_{k|k}^{1/2}$ is already available in the MCC-KF algorithm, the estimator implementation implies its re-calculation by using the so-called Joseph stabilized equation for updating the error covariance matrix $P_{k|k}$ at the last line of Algorithm~1. To distinguish the matrix $P_{k|k}$ used in the gain $K_k$ calculation by formula~\eqref{mcc:K:eq2} from the matrix obtained from the Joseph stabilized equation, we use notation $\hat P_{k|k}$ and $P_{k|k}$ for these two cases, respectively. It is worth noting here that the Joseph stabilized formula ensures the symmetric form of error covariance matrix $P_{k|k}$ and, hence, it
is recognized to be the preferable implementation strategy of any conventional filtering algorithm. In a similar way, we factorize
\begin{align*}
AA^{\top} & = (I\!- K_{k}H_k)P_{k|k-1}(I\!- K_{k}H_k)^{\top}\!\!+K_k R_k K_k^{\top} \\
& =[(I\!- K_{k}H_k)P_{k|k-1}^{1/2}, \; K_kR_k^{1/2}] \\
& \times [(I\!- K_{k}H_k)P_{k|k-1}^{1/2}, \; K_kR_k^{1/2}]^{\top} \\
& = RR^{\top}  = P_{k|k} = [P_{k|k}^{1/2}, \; 0][P_{k|k}^{1/2}, \; 0]^{\top}.
\end{align*}

Having summarized the formulas above, we obtain the following Cholesky-based square-root MCC-KF method.
%_______________________________________
\begin{codebox}
\Procname{{\bf Algorithm 1a}. $\proc{SR MCC-KF}$ ({\it Cholesky-based MCC-KF})}
\zi \textsc{Initialization:}($k=0$)
\zi \>Apply Cholesky factorization: $\Pi_0 = \Pi_0^{1/2}\Pi_0^{{\top}/2}$;
\zi \>Set initial values: $\hat x_{0|0} = \bar x_0$, $P_{0|0}^{1/2} = \Pi_0^{1/2}$;
\zi \textsc{Time Update}: ($k=\overline{1,N}$)
\li \>Compute $\hat x_{k|k-1}$ in line~\ref{mcc:p:X} of Algorithm~1;
\li \>Build the pre-array and lower triangularize it \label{mcc:sr1:p:P}
\zi \>$\underbrace{
\begin{bmatrix}
F_{k-1}P_{k-1|k-1}^{1/2}, \; G_{k-1}Q_{k-1}^{1/2}
\end{bmatrix}
}_{\mbox{\scriptsize Pre-array $\mathbb A$}} {\mathbb Q}_1
   =
\underbrace{
\begin{bmatrix}
P_{k|k-1}^{1/2}, \;  0
\end{bmatrix}
}_{\mbox{\scriptsize Post-array $\mathbb R$}}$;
\zi \>Read-off from post-array the factor $P_{k|k-1}^{1/2}$;
\zi \textsc{Measurement Update}: ($k=\overline{1,N}$)
\li \>Compute $\lambda_k$ by formula~\eqref{eq:lambda};
\li \>Build the pre-array and lower triangularize it \label{mcc:sr1:f:K}
\zi \>$\underbrace{
\begin{bmatrix}
P_{k|k-1}^{-{\top}/2}, \; \lambda_k^{1/2}H_k^{\top}R_k^{-{\top}/2}
\end{bmatrix}
}_{\mbox{\scriptsize Pre-array $\mathbb A$}} {\mathbb Q}_2
   =
\underbrace{
\begin{bmatrix}
\hat P_{k|k}^{-{\top}/2}, \;  0
\end{bmatrix}
}_{\mbox{\scriptsize Post-array $\mathbb R$}}$;
\zi \>Read-off from post-array the factor $[\hat P_{k|k}^{-{\top}/2}]$;
\li \>Compute $K_{k}  = \lambda_k [\hat P_{k|k}^{-{\top}/2}]^{-{\top}}[\hat P_{k|k}^{-{\top}/2}]^{-1}H_k^{\top} R_k^{-1}$;
\li \>Calculate the state $\hat x_{k|k}$ in line~\ref{mcc:f:X} of Algorithm~1;
\li \>Build the pre-array and lower triangularize it  \label{mcc:sr1:f:P}
\zi \>$\underbrace{
\begin{bmatrix}
(I\!- K_{k}H_k)P_{k|k-1}^{1/2}, \; K_kR_k^{1/2}
\end{bmatrix}
}_{\mbox{\scriptsize Pre-array $\mathbb A$}} {\mathbb Q}_3
   =
\underbrace{
\begin{bmatrix}
P_{k|k}^{1/2}, \; 0
\end{bmatrix}
}_{\mbox{\scriptsize Post-array $\mathbb R$}}$;
\zi \>Read-off from post-array the factor $P_{k|k}^{1/2}$.
\end{codebox}
%\setlinenumberplus{mcc:sr:p:X}{1}

It is not difficult to see that the numerical behaviour of Algorithm~1a heavily depends on a condition number of matrices $P_{k|k-1}^{1/2}  \in {\mathbb R}^{n\times n}$ and $\hat P_{k|k}^{1/2}  \in {\mathbb R}^{n\times n}$ because their inversion is required in Algorithm~1a. The method can be improved by avoiding the matrix inversion operation.

We construct an alternative robust Cholesky-based variant by using equation~\eqref{mcc:K:eq1} for computing the gain matrix $K_k$. As a result, our new square-root method requires the inverse of the lower triangular matrix $R_{e,k}^{1/2} \in {\mathbb R}^{m\times m}$, only. Indeed, we factorize equation $R_{e,k} = \lambda_k H_kP_{k|k-1}H^{\top} + R_k$ in a similar way as shown above and, then, utilize a stable orthogonal transformation for updating the resulted square-root factors, i.e. we get
\begin{align*}
AA^{\top} & = \lambda_k H_kP_{k|k-1}H^{\top} + R_k \\
& =[\lambda_k^{1/2} H_kP_{k|k-1}^{1/2}, \; R_k^{1/2}][\lambda_k^{1/2} H_kP_{k|k-1}^{1/2}, \; R_k^{1/2}]^{\top} \\
& = RR^{\top}  = R_{e,k} = [R_{e,k}^{1/2}, \; 0][R_{e,k}^{1/2}, \; 0]^{\top},
\end{align*}
and the gain matrix $K_k$ is then calculated by equation~\eqref{mcc:K:eq1} by using the available value $[R_{e,k}^{1/2}]$ as follows:
\[
K_{k} = \lambda_k P_{k|k-1}H_k^{\top}R_{e,k}^{-1} =  \lambda_k P_{k|k-1}H_k^{\top}[R_{e,k}^{1/2}]^{-{\top}}[R_{e,k}^{1/2}]^{-1}.
\]

Thus, we summarize an alternative Cholesky-based implementation for the MCC-KF estimator.
%_______________________________________
\begin{codebox}
\Procname{{\bf Algorithm 1b}. $\proc{SR MCC-KF}$}
\zi \textsc{Initialization:} Repeat from Algorithm 1a.
\zi \textsc{Time Update}: Repeat from Algorithm 1a.
\zi \textsc{Measurement Update}: ($k=\overline{1,N}$)
\li \>Compute $\lambda_k$ by formula~\eqref{eq:lambda};
\li \>Build the pre-array and lower triangularize it \label{mcc:sr1:f:K}
\zi \>$\underbrace{
\begin{bmatrix}
\lambda_k^{1/2} H_kP_{k|k-1}^{1/2}, \; R_k^{1/2}
\end{bmatrix}
}_{\mbox{\scriptsize Pre-array $\mathbb A$}} {\mathbb Q}_2
   =
\underbrace{
\begin{bmatrix}
R_{e,k}^{1/2}, \;  0
\end{bmatrix}
}_{\mbox{\scriptsize Post-array $\mathbb R$}}$;
\zi \>Read-off from post-array the factor $[R_{e,k}^{1/2}]$;
\li \>Compute $K_{k}  = \lambda_k P_{k|k-1}H_k^{\top}[R_{e,k}^{1/2}]^{-{\top}}[R_{e,k}^{1/2}]^{-1}$;
\li \>Compute $\hat x_{k|k}$ by formula in line~\ref{mcc:f:X} of Algorithm~1;
\li \>Use formula in line~\ref{mcc:sr1:f:P} of Algorithm~1a to find $P_{k|k}^{1/2}$.
\end{codebox}
%\setlinenumberplus{mcc:sr:p:X}{1}

As can be seen, the suggested  square-root Algorithm~2b  requires the innovation covariance matrix $R_{e,k}^{1/2}$ inversion, only. Thus, it is expected to possess a better numerical behavior compared to the conventional implementation in Algorithm~1 and the Cholesky-based Algorithm~1a. Next section provides the results of numerical experiments.

\section{NUMERICAL EXPERIMENTS}

We first wish to justify a theoretical derivation of the presented square-root MCC-KF Algorithm~1b.

\begin{exmp} \label{ex:1} In the Radar Tracking Example from~\cite[p.~227]{2015:Grewal:book}, the signals are processed by
the filter in order to determine the position of maneuvering airborne objects. The system state is defined as follows: $x_k = [r_k, \: \dot{r}_k, \: U_k^1, \: \theta_k, \: \dot{\theta}_k, \: U^2_k]^{\top}$ where $r_k$ is the range of the vehicle at time $t_k$, $\dot{r}_k$ is the range rate of the vehicle at time $t_k$, $U_k^1$ is the maneuvering-correlated state noise, $\theta_k$ is the bearing of the vehicle at time $t_k$, $\dot{\theta}_k$ is the bearing rate of the vehicle at time $t_k$, $U^2_k$ is the maneuvering-correlated state noise. The model dynamics is given as follows:
\begin{align*}
x_{k} & =
\begin{bmatrix}
1 & T & 0 & 0 & 0 & 0 \\
0 & 1 & 1 & 0 & 0 & 0 \\
0 & 0 & \rho & 0 & 0 & 0 \\
0 & 0 & 0 & 1 & T & 0 \\
0 & 0 & 0 & 0 & 1 & 1 \\
0 & 0 & 0 & 0 & 0 & \rho
\end{bmatrix}
x_{k-1} +
\begin{bmatrix}
0  \\
0  \\
w_{k-1}^1  \\
0  \\
0  \\
w_{k-1}^2
\end{bmatrix},
\end{align*}
 where $\rho = 0.5$ is correlation coefficient and $T = 10$ is the sampling period in seconds. The measurements are  provided by
\begin{align*}
y_k & =
\begin{bmatrix}
1 & 0 & 0 & 0 & 0 & 0 \\
0 & 0 & 0 & 1 & 0 & 0
\end{bmatrix}
x_k +
\begin{bmatrix}
v_k^1 \\
v_k^2
\end{bmatrix}
\end{align*}
The MCC-KF estimators are tested in the presence of, the so-called, impulsive noise scenario, which is used for simulating the outliers in the examined Gaussian state-space model:
\begin{align*}
w_k  & \sim {\cal N}(0, Q)+\mbox{\tt Shot noise}  \mbox{(20\% are corrupted)},\\
v_k  & \sim {\cal N}(0, R)+\mbox{\tt Shot noise}   \mbox{(20\% are corrupted)}
\end{align*}
where the covariance matrices $Q$ and $R$ are given by
\[
Q =
\begin{bmatrix}
0 & 0 & 0 & 0 & 0 & 0 \\
0 & 0 & 0 & 0 & 0 & 0 \\
0 & 0 & \sigma_1^2 & 0 & 0 & 0 \\
0 & 0 & 0 & 0 & 0 & 0 \\
0 & 0 & 0 & 0 & 0 & 0 \\
0 & 0 & 0 & 0 & 0 & \sigma_2^2
\end{bmatrix} \quad  \mbox{and} \quad
R =
\begin{bmatrix}
\sigma_{r}^2 & 0 \\
0 & \sigma_{\theta}^2
\end{bmatrix}
\]
with $\sigma_{r}^2 = (1000\mbox{ m})^2$, $\sigma_{\theta}^2 = (0.017\mbox{ rad})^2$, $\sigma_1^2 = (103/3)^2$ and $\sigma_2^2 = 1.3 \times 10^{-8}$. Finally, the initial values for each estimator to be examined are the following: $x_0 \sim {\cal N}(\bar x_0, \Pi_0)$ where $\bar x_0 = 0$ and
\[
\Pi_0 =
\begin{bmatrix}
            \sigma_r^2&     \displaystyle\frac{\sigma_r^2}{T}&              0&        0&               0& 0 \\
        \displaystyle\frac{\sigma_r^2}{T} & \displaystyle\frac{2\sigma_r^2}{T^2}+\sigma_1^2& 0&        0&               0& 0  \\
                    0&                          0& \sigma_1^2& 0&               0& 0  \\
                    0&                          0&        0& \sigma_{\theta}&   \displaystyle\frac{\sigma_{\theta}}{T}& 0  \\
                    0&                          0&        0& \displaystyle\frac{\sigma_{\theta}}{T}& \displaystyle\frac{2\sigma_{\theta}}{T^2}+\sigma_1^2& 0  \\
                    0&                          0&        0&               0&                              0&\sigma_2^2
\end{bmatrix}.
\]
\end{exmp}

To simulate the impulsive noise (the shot noise), we follow the approach suggested in~\cite{2016:Izanloo}. The corresponding Matlab routine \verb"Shot_noise" is also published recently in~\cite{2020:AJC:Kulikova}. The magnitude of each impulse is chosen randomly from the uniform discrete distribution in the interval $[0,5]$. The time instances where the outliers occur are also chosen randomly on the interval $t_k \in [21, 300]$ by using the uniform distribution. When the `exact' solution $x_{exact}(t_k)$ and the measurements $y_k = y(t_k)$ are simulated on the examined interval $t_k \in [1, 300]$, we solve the inverse problem, i.e. the dynamic state is to be estimated from the observed signal $y(t_k)$ by various filtering algorithms. They are all tested within equal conditions, i.e. they utilize the same filters' initials, the same measurements and the same noise covariances. Finally, the experiment is repeated for $M=100$ times and the root mean square error (RMSE) is calculated in these Monte Carlo runs.

\begin{figure}
\includegraphics[width=0.5\textwidth]{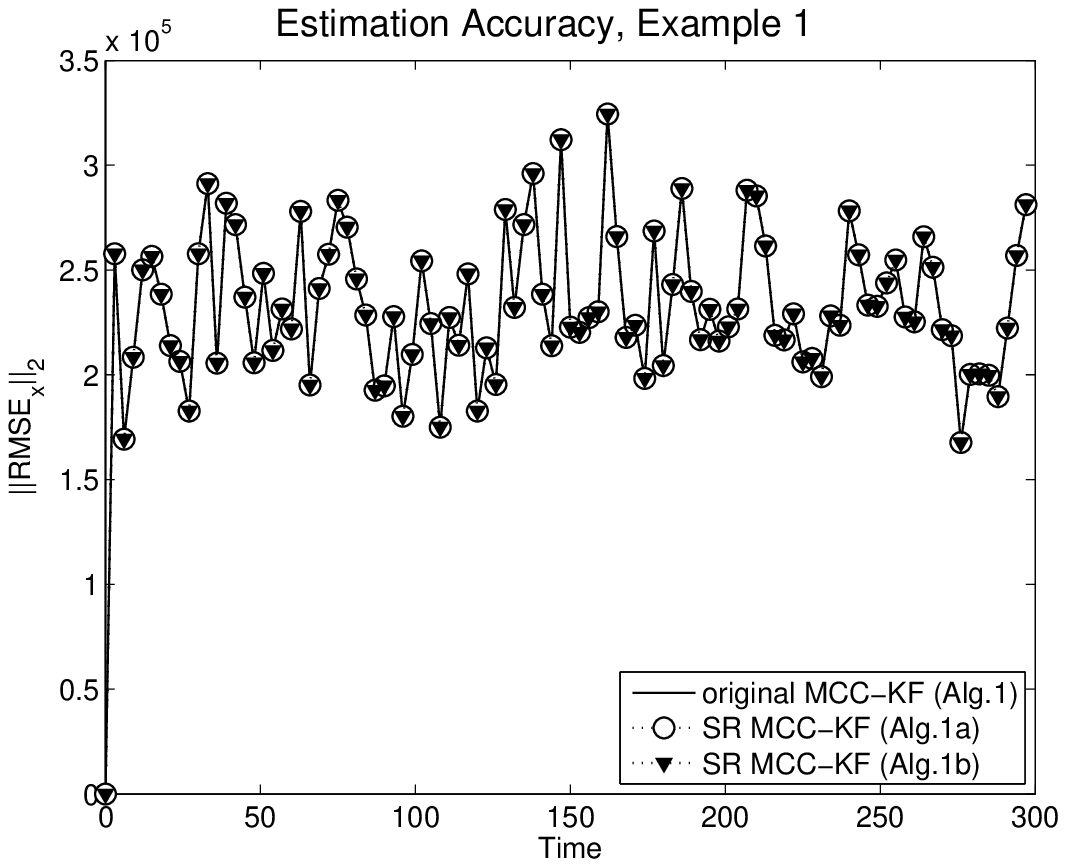}
\caption{The total $\|\mbox{\rm RMSE}_{x_i}(t_k)\|_2$  ($i=1, \ldots, 6$) calculated for each MCC-KF estimator in case of radar tracking scenario in Example~\ref{ex:1}.}  \label{fig:1}
\end{figure}

Fig.~\ref{fig:1} illustrates the total RMSE in all six components of the dynamic state averaged over $M=100$ Monte Carlo runs. As can be seen, all implementation methods produce the same result, i.e. they work with the same estimation accuracy. This substantiates an algebraic equivalence of all MCC-KF algorithms under examination and the correctness of our theoretical derivations presented in Section~\ref{main:result}. Meanwhile, the difference in numerical behaviour of the conventional Algorithm~1, the previously suggested square-root Algorithm~1a and the newly derived square-root Algorithm~1b can be observed in ill-conditioned state estimation scenario. For that, we consider Example~\ref{ex:2}.

%______________________________________________________________________

\begin{exmp} \label{ex:2}
Consider the radar tracking problem in Example~\ref{ex:1} with Gaussian uncertainties only, where the dynamic state is observed through the following measurement scheme:
\begin{align*}
y_k & =
\begin{bmatrix}
1 & 1 & 1 & 1 & 1 &  1\\
1 & 1 & 1 & 1 & 1 &  1 +\delta
\end{bmatrix}
x_k +
\begin{bmatrix}
v_k^1 \\
v_k^2
\end{bmatrix}
\end{align*}
where $x_0 \sim {\cal N}(0, I_6)$ and $v_k \sim {\cal N}(0,\delta^2 I_2)$. The parameter $\delta$ is used for simulating roundoff effect, i.e. it is assumed to be $\delta^2<\epsilon_{roundoff}$, but $\delta>\epsilon_{roundoff}$ and $\epsilon_{roundoff}$ stands for the unit roundoff error.
\end{exmp}

Measurement scheme in Example~\ref{ex:2} allows for simulating various ill-conditioned scenarios including the continuous-discrete estimation methods as discussed in~\cite{2017:Kulikov:IET,2018:Kulikov:IET,2019:Kulikov:IJRNC}. More precisely, when the ill-conditioning parameter $\delta$ tends to machine precision limit, $\delta \to \epsilon_{roundoff}$, we observe a degradation of the underlying Riccati-type recursion. The discussed numerical instability of the conventional filtering methods is arisen from the matrix inversion $R_{e,k}$ that becomes almost singular after a few filtering steps; see the third reason of the KF divergence in~\cite[p.~288]{2015:Grewal:book}. Within the square-root implementation approach, the inverse of its triangular Cholesky factor is required instead.

\begin{figure}
\includegraphics[width=0.5\textwidth]{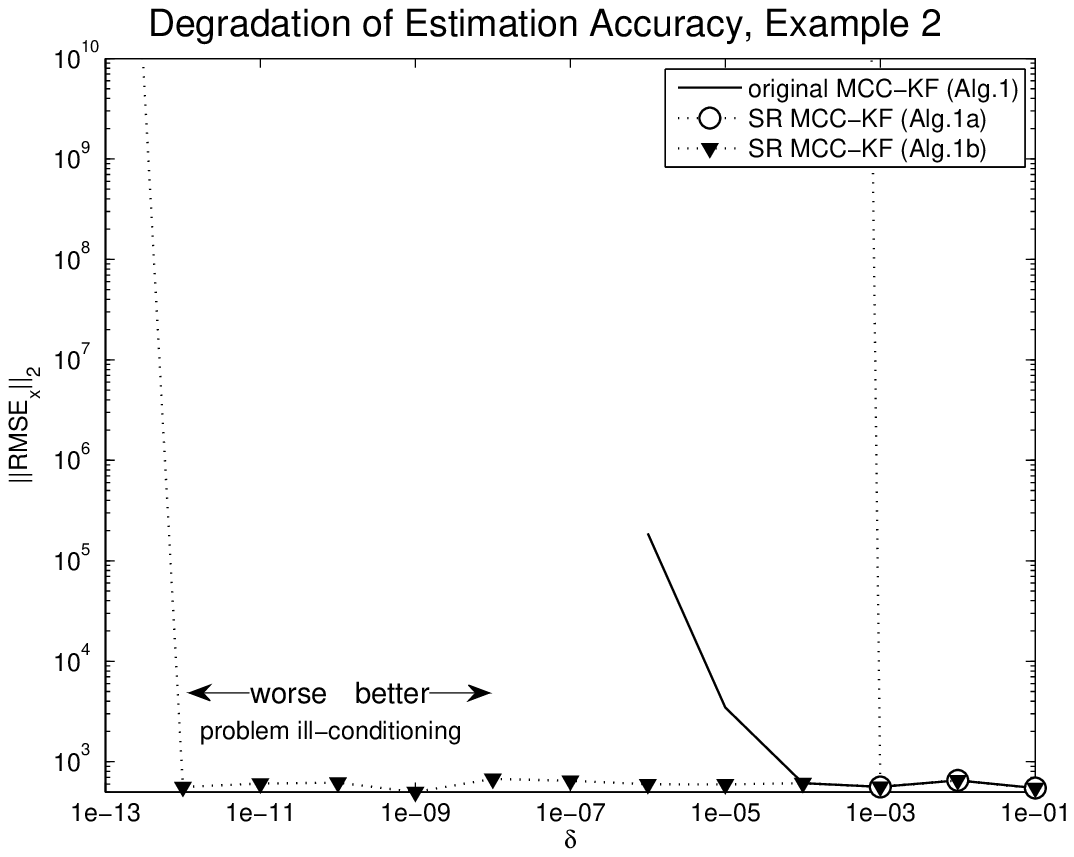}
\caption{The degradation of total RMSE for each MCC-KF estimator under examination in case of ill-conditioned radar tracking cases in Example~\ref{ex:2}.}  \label{fig:2}
\end{figure}

Fig.~\ref{fig:2} illustrates a degradation of the resulted estimation accuracies when  $\delta \to \epsilon_{roundoff}$ for each MCC-KF implementation under examination. As it was anticipated the square-root MCC-KF Algorithm~1b is the most robust implementation method among three algorithms examined. We observe that all MCC-KF methods work equally accurate and with small estimation errors in well-conditioned scenarios, i.e. when $\delta$ is large, which corresponds to ill-conditioned parameter  $\delta = 10^{-1}$ and $\delta = 10^{-2}$. However, while the problem ill-conditioning increases the conventional Algorithm~1 and the square-root Algorithm~1a fail to solve the stated problem. Indeed, for $\delta \le 10^{-4}$ they both provide either a very large total RMSE or yield \verb"NaN" (it states for `Not a Number' in MATLAB), which is not plotted. It is worth noting here that the conventional MCC-KF implementation in Algorithm~1 degrades even a bit slower than the previously published square-root Algorithm~1a. A possible explanation for such behaviour is that the conventional MCC-KF implementation implies the  Joseph  stabilized equation for the error covariance matrix calculation. Indeed, the MCC-KF estimator in Algorithm~1 fails at $\delta = 10^{-5}$ because the resulted accuracies are very large, i.e. it means no correct digits in the obtained estimates. Meanwhile, the square-root Algorithm~1a fails a bit faster, i.e. at $\delta = 10^{-4}$. Finally, the newly-derived square-root method in Algorithm~1b is the most stable implementation method. It works  accurately until the ill-conditioned parameter $\delta = 10^{-13}$, i.e. it is able to manage the ill-conditioned scenarios of Example~\ref{ex:2}. This is the only one robust Cholesky-based implementation existed for the MCC-KF estimator in engineering literature so far.

\section{CONCLUSION}\label{sect5}

In this paper, the problem of designing the numerically stable square-root methods for the maximum correntropy criterion Kalman filter is discussed. The first robust (with respect to roundoff errors) square-root MCC-KF algorithm has been found within the class of Cholesky factorization-based implementations. Nowadays, this is the only one reliable Cholesky-based method existed for the MCC-KF estimator in engineering literature. The square-root solution is proposed for the MCC-KF filtering for a case of the scalar adjusting parameter involved. The derivation of a square-root solution within the Cholesky decomposition for the MCC-KF methods with matrix-type adjusting weights involved is an open question for a future research.

%\bibliography{ifacconf}             % bib file to produce the bibliography
                                     % with bibtex (preferred)
%\bibliographystyle{ifacconf-harvard}

\bibliographystyle{ifacconf}
\bibliography{BibTex_Library/books,%
              BibTex_Library/KFDistribution_Robust,%
              BibTex_Library/KF_Chandrasekhar,%
              BibTex_Library/KFMCC_Riccati,%
              BibTex_Library/KFMCC_Applications,%
              BibTex_Library/KFDiff_Chandrasekhar,%
              BibTex_Library/KF_Riccati,%
              BibTex_Library/CKF_Riccati,%
              BibTex_Library/UKF_Riccati,%
              BibTex_Library/EKF_all,%
              BibTex_Library/KF_Applications,%
              BibTex_Library/Lin_Algebra}

\end{document}